\documentclass[%
 reprint,
 nofootinbib,
 amsmath,amssymb,
 aps,
 prd,
]{revtex4-2}

\usepackage{aas_macros}
\usepackage{booktabs}   
\usepackage{multirow}   
\usepackage{siunitx}    
\usepackage{orcidlink}
\usepackage{graphicx}
\usepackage{dcolumn}
\usepackage{bm}
\usepackage{float}          

\makeatletter
\let\newfloat\newfloat@ltx  
\makeatother

\usepackage{algorithm}
\usepackage{graphicx}        
\usepackage{enumitem}
\usepackage{microtype}
\usepackage{xcolor}
\usepackage{xspace}
\usepackage{changes}
\renewcommand{\added}[2][]{#2}
\renewcommand{\deleted}[2][]{}

\renewcommand{\comment}[2][]{}

\definecolor{myhyperlinkcolor}{HTML}{C31E24}
\hypersetup{
    colorlinks=true,
    citecolor=myhyperlinkcolor,
    linkcolor=myhyperlinkcolor,
    urlcolor=myhyperlinkcolor
}

\newtheorem{definition}{Definition}

\newcommand{\project}[1]{\textsf{#1}\xspace}

\newcommand{\morphz}{\project{MorphZ}}

\newcommand{\bilby}{\project{bilby}}
\newcommand{\dynesty}{\project{dynesty}}
\newcommand{\bilbymcmc}{\project{bilby-MCMC}}

\newcommand{\lnZns}{\ensuremath{\log(\hat{z})_{\mathrm{NS}}}}

\begin{document}

\preprint{APS/123-QED}

\title{Enhancing evidence estimation through informed probability density approximation}


\newcommand{\AUTMaths}{Department of Mathematical Sciences, Auckland University of Technology, Private Bag 92006, Auckland 1142, New Zealand}
\newcommand{\Manly}{Manly Astrophysics, 15/41-42 East Esplanade, Manly, NSW 2095, Australia}
\newcommand{\UoAStats}{Department of Statistics, University of Auckland, 38 Princes St, Auckland, New Zealand}

\author{El Mehdi Zahraoui$^{1}$\orcidlink{0009-0006-0900-3824}}
\author{Patricio Maturana-Russel$^{1,2}$\orcidlink{0000-0002-5211-9818}}
\author{Avi Vajpeyi$^{2}$\orcidlink{0000-0002-4146-1132}} 
\author{Willem van Straten$^{3}$\orcidlink{0000-0003-2519-7375}}
\author{Renate Meyer$^{2}$\orcidlink{0000-0003-0268-8569}}
\author{Sergei Gulyaev$^{1}$\orcidlink{0000-0003-0186-5551}}

\affiliation{$^{1}$\AUTMaths}
\affiliation{$^{2}$\UoAStats}%
\affiliation{$^{3}$\Manly}

\date{\today}

\begin{abstract}
We introduce the \emph{Morph} approximation, a class of product approximations of probability densities that selects low–order disjoint parameter blocks by maximizing the sum of their total correlations. We use the posterior approximation via Morph as the importance distribution in optimal bridge sampling. We denote this procedure by \morphz, which serves as a post–processing estimator of the marginal likelihood.
The \morphz estimator requires only posterior samples \deleted{together with the prior and likelihood}, and is fully agnostic \deleted{to} \added{regarding} the choice of sampler.
We evaluate \morphz's performance across statistical benchmarks, pulsar timing array (PTA) models, compact binary coalescence (CBC) gravitational-wave (GW) simulations and the GW150914 event. 
Across these applications, spanning low to high dimensionalities, \morphz yields accurate evidence \added{estimates} at substantially reduced computational cost relative to standard approaches\deleted{, and can improve these estimates even when posterior coverage is incomplete}. \added{We have found that when these approaches fail to provide accurate estimates, \morphz has proven to either resolve the estimation failure or significantly improve the results.} Its bridge sampling relative error diagnostic provides conservative uncertainty estimates.
Because \morphz operates directly on posterior draws, it complements exploration-oriented samplers by enabling fast and reliable evidence estimation, while it can be seamlessly integrated into existing inference workflows.
\end{abstract}

\maketitle


\section{Introduction}

Across modern astrophysics and cosmology, Bayesian inference has become foundational for both parameter estimation and hypothesis testing. In this framework, prior beliefs about model parameters are updated to posterior beliefs as new data are acquired, providing a coherent probabilistic framework for quantifying uncertainty in both the parameters and the resulting model predictions.  A key quantity is the Bayesian evidence (marginal likelihood), which underpins probabilistic model comparison and hypothesis assessment. Thanks to its ability to combine heterogeneous datasets and incorporate prior information consistently, the Bayesian framework is well suited to experiments across a wide range of scales, and is particularly advantageous for decade-spanning observational programs. 

Among these experiments, the pulsar timing array (PTA) stands out as a well-suited experiment that employs Bayesian analysis techniques to update the status of gravitational-wave background detection. Accordingly, \added{several of} the latest PTA data releases provide Bayesian evidence favoring the presence of a gravitational-wave background (GWB) in their datasets~\cite{reardon2023search,agazie2023nanograv,antoniadis2023second}. These data releases are a product of Herculean efforts involving observations of hundreds of millisecond pulsars (MSPs), detailed per-pulsar noise characterization~\cite{agazie2023nanograv,antoniadis2023second,zic2023parkes}, and joint analysis aimed at GWB detection~\cite{reardon2023search,agazie2023nanograv,antoniadis2023second,miles2023meerkat}. Currently, the PTA model selection and evidence estimation \deleted{are carried out through} \added{are performed using} different techniques \citep{agazie2023nanograv,antoniadis2023second}: thermodynamic integration \citep[]{lartillot2006computing}, nested sampling \citep{skilling2006nested}, model reweighting \citep{hourihane2023accurate}, Savage-Dickey density ratio approximation \citep{carlin1995bayesian,lodewyckx2011tutorial}. 
The efficiency and accuracy of evidence estimation methods employed by the PTA are key for robust detection claims and lower computational costs. The future Square Kilometre Array (SKA) \cite{smits2009pulsar,stappers2018prospects}  dataset will significantly increase the complexity of hierarchical noise and signal models, making efficient evidence estimation a practical necessity to keep the computational cost tractable. 

In parallel with pulsar timing arrays, ground-based gravitational-wave astronomy provides a compelling setting in which Bayesian model selection plays a key role in astrophysical inference. 
Within the LIGO--Virgo--KAGRA (LVK) collaboration, the Bayesian evidence plays a central role in adjudicating between competing waveform models, ranking candidate events, identifying incoherent noise transients, and performing tests of general relativity~\cite{2024PhRvD.109b3019P,2025arXiv250421147R,2022MNRAS.516.5309V}. 
Recent analyses have shown that subtle differences in waveform physics, such as the inclusion of spin precession or higher-order modes, can lead to biased scientific conclusions unless model choices are explicitly validated through evidence-based comparison~\cite{2024PhRvD.109b3019P,2025arXiv250421147R}.
\added{In addition, the evidence can also be used for waveform model averaging, weighting each waveform family in proportion to its support to marginalize over waveform systematics rather than relying on a single preferred model \citep{2020PhRvD.101f4037A}.}
As the complexity of 
Compact Binary Coalescence (CBC) signal hypotheses continues to grow, LVK inference pipelines increasingly rely on accurate and scalable evidence estimation. 
This need will intensify for next-generation instruments, including the Einstein Telescope~\cite{ET2010}, Cosmic Explorer~\cite{CE2019}, and LISA~\cite{LISA2017}, which will produce significantly larger datasets and more intricate likelihoods. 
Efficient and reliable computation of evidence is therefore essential for  model selection and hypothesis testing across current and future gravitational-wave observatories.

The Bayesian evidence, commonly known as the \textit{marginal likelihood}, is a multidimensional integral over model parameters that is generally estimated numerically. The marginal likelihood literature provides a wide variety of approaches,  spanning analytic approximations, sampling-based estimators~\cite{chib1995understanding,meng1996simulating,skilling2006nested}, and path-based techniques~\cite{friel2008marginal,xie2011improving}. Each newly proposed method aims to lower the computational expense of marginal likelihood estimation while improving estimator accuracy in challenging Bayesian models. Among these methods, several accurate evidence estimators rely on importance sampling distributions tuned to the posterior samples, such as bridge sampling (BS; \cite{meng1996simulating,gronau2017tutorial}), nested importance sampling (NIS;~\cite{chopin2010properties,feroz2013importance}), Generalized Annealing importance sampling (GAIS;~\cite{russel2017bayesian}), and Generalized Steppingstone Sampling (GSS;~\cite{fan2011choosing}).  \citet{chan2015marginal} show that an optimal importance sampling distribution is the distribution that minimizes the cross-entropy (equivalently, the \textit{Kullback–Leibler} divergence) relative to the posterior. In practice, selecting an importance distribution on a case-by-case basis limits its practicality for routine estimation. Common heuristics include modeling the posterior as a product of marginal distributions fitted to posterior samples or a multivariate Gaussian with a covariance matrix estimated from those samples~\cite{fan2011choosing,chan2015marginal}. 
These heuristics limit the importance sampling distribution to either preserve the marginal shapes or to assume Gaussian marginals while capturing linear correlations. 
Alternatively, better importance distributions employ different types of copulas tuned to posterior samples \cite{silva2008copula,dellaportas2019importance,kauermann2014penalized,Craiu20115,Schmidl20131,South2019753}.
These copula-based importance densities capture skewness and tail dependence better than a multivariate normal and are usually cheap to fit to posterior samples. However, they were only tested on moderate-dimensional regression and random-effects models~\cite{silva2008copula}, and need to be further tested on high-dimensional and multimodal problems. Additionally, training a normalizing flow on the posterior samples has proven to be effective and more flexible as an importance distribution for up to $21$ dimensions~\cite{polanska2024learned}; however, scalability to higher dimensionalities and more complex posteriors remains to be systematically assessed.

The marginal likelihood computational cost decreases as the proposal more closely matches the posterior~\cite{chan2015marginal}. Accordingly, we present a novel approach based on a class of product approximations to probability densities that more fully exploits available posterior information. This approximation enhances marginal likelihood estimators making evidence available at a fraction of the cost relative to popular approaches in astrophysics. This new family of product approximations is built on the notion of \textit{total correlation} \cite{watanabe1960information}, an information-theoretic measure that captures both linear and non-linear dependencies; we refer to it as \textit{Morph} approximation. By reducing a high-dimensional posterior to a product of low-dimensional factors,  the \textit{Morph} approximation achieves scalable inference and strong performance from low to high dimensionalities. To construct the \textit{Morph} approximation, we use standard kernel density estimation (KDE) algorithms \cite{virtanen2020scipy,pauli_virtanen_2020_4406806}, yielding an approximation that preserves empirical marginal shapes and captures the dominant dependencies at low cost. In combination with bridge sampling, the \morphz estimator delivers efficient and accurate marginal likelihood estimates, outperforming commonly used  methods on challenging problems and remaining effective for high-dimensional and expensive-likelihood settings. This approach prioritizes the posterior coverage; once coverage is sufficient, the evidence can be computed at minimal additional cost.

The remainder of this paper is organized as follows. In Section~\ref{evidence}, we introduce the marginal likelihood and the corresponding estimation methods. Section~\ref{method} defines the Morph approximation, the \morphz estimator, and describes the procedure used to construct this estimator. Section~\ref{appli:stats} evaluates the estimator's performance on a set of challenging statistical benchmarks. Sections~\ref{appli:pta} and~\ref{appli:LVK} demonstrate the efficiency and accuracy of the \morphz estimator through applications to pulsar timing array and LIGO-VIRGO-KAGRA (LVK) gravitational wave analyses. In Section~\ref{discuss}, we address practical aspects of the Morph approximation and bridge sampling. Finally, Section~\ref{conclusion} concludes with an assessment of the effectiveness of \morphz and outlines future directions.

\section{Marginal likelihood estimation}
\label{evidence}

The \textit{marginal likelihood} or evidence is defined by \textit{Bayes'} rule as
\begin{equation}    \textit{z}(X|M)=\int_{\Theta}L(X|\boldsymbol\theta,M)\pi(\boldsymbol\theta|M)\text{d}\boldsymbol\theta,
    \label{marginal_like}
\end{equation}
where $L(X|\boldsymbol\theta,M)$ is the likelihood function of the model $M$ over the dataset $X$, and $\pi(\boldsymbol\theta|M)$ is a proper prior density of the parameter vector $\boldsymbol{\theta}$. The marginal likelihood $z(X|M)$, or simply $z$ from here onward,  is a quantity routinely used to measure the quality of the Bayesian fit. 
If two models, $M_0$ and $M_1$, fitted to a dataset $X$, are assumed to be \deleted{a $priori$} equally probable, then the ratio of their marginal likelihoods \deleted{is equal to the \textit{Bayes factor},} \added{is defined as the \textit{Bayes factor},}
\begin{equation}
    \text{BF}_{(M_{0} / M_{1})}=\frac{z(X|M_{0})}{z(X|M_{1})}.
    \label{bayes_factor}
\end{equation}
 This Bayesian criterion selects, between $M_{0}$ and $M_{1}$, the model best supported by the observed data $X$. The Bayes factor embodies  \textit{Occam's razor}, discouraging excessive complexity and mitigating overfitting.
 The marginal likelihood can be evaluated analytically only for special choices of priors and likelihoods (conjugate prior-likelihood pairs) whose product integrates in closed form. In general, the multi-dimensional integral is intractable and must be approximated numerically. 
To this end, multiple Markov chain Monte Carlo (MCMC) methods have been proposed to estimate the evidence, such as the harmonic mean~\citep{newton1994approximate} 
\added{, importance sampling \citep{kloek1978bayesian}}, nested sampling~\citep{skilling2006nested}, path sampling~\citep{gelman1998simulating}, and steppingstone sampling~\citep{xie2011improving}. In the subsequent literature, an importance density fitted to posterior draws has been introduced as an extension of these methods to increase the accuracy of the marginal-likelihood estimates \added{(bridge sampling \cite{gelman1998simulating}, nested importance sampling \cite{chopin2010properties}, generalized steppingstone sampling \cite{fan2011choosing}).}
These extension methods have proven to reduce the computational cost while being more accurate than the original MCMC methods. 

In this work, we introduce the \textit{Morph approximation}: an information-based product approximation of the posterior density that can serve as an optimal importance proposal. We combine the Morph approximation with optimal bridge sampling~\cite{gelman1998simulating,gronau2017tutorial}, which yields an automated and highly efficient estimator  for the marginal likelihood (\morphz) that performs well in typical applications and demonstrates improved performance in statistically challenging scenarios (see Section~\ref{appli:stats}).

\section{Method}
\label{method}

We begin by introducing the formal mathematical definitions underlying the Morph approximation. In this framework, we consider a probability density over a set of random variables (which, in Bayesian applications, corresponds to the posterior distribution). The Morph approximation constructs a tractable, normalized probability density from low-order factors while retaining its most important dependence structure. This is achieved by decomposing the full probability density into a product of lower-dimensional disjoint factors, each defined on a block of variables of equal length. These blocks are formed from groups of variables that are strongly correlated, and are chosen so as to maximize the total correlation captured within the blocks. The resulting low-dimensional factors can then be estimated using standard statistical techniques from available samples and combined to form an efficient proposal distribution for subsequent bridge-sampling evidence estimation.

\subsection{Morph approximation}
\label{subsec:morph}
Let $P(\boldsymbol{\theta})$ be a $d$-dimensional joint probability distribution with $\boldsymbol{\theta}=(\theta_1,\ldots,\theta_d)^\top\in\mathbb{R}^d$ the parameter vector. A product approximation of $P(\boldsymbol{\theta})$ is defined as a product of lower-order distributions, such that the product is a probability extension of these lower-order distributions that satisfies the unity sum property~\cite{lewis1959approximating}.\\
\begin{definition}\label{def:morph approx.}
Let $\Gamma = \{1,\dots,d\}$ represent the set of integers from 1 to $d$, fix $L\in\Gamma$ by choosing one element from $\Gamma$ , then \( m = \left\lfloor d/L \right\rfloor \in \mathbb{N}^{*} \) is the integer number of times that $L$ divides $d$, with $\mathbb{N}^{*}$ the natural number set excluding zero. Let 
\begin{equation}
\begin{split}
    \mathcal{F}_{L} &= \{\mathcal{B}_{L}=\{b_{1},\dots,b_{m} \}:b_{i}\subseteq\Gamma,|b_{i}|=L,\\
    &b_{i}\cap b_{j}=\varnothing\,\,\,\forall i \neq j  \},
\end{split}
\label{group_family}
\end{equation}
with $\mathcal{B}_{L}$ the partial partition that includes all non-intersecting blocks $b_{i}$ of  $\Gamma$ containing $L$ elements. For $\mathcal{B}_{L}\in \mathcal{F}_{L}$, let the singletons residual index set $\mathcal{S}$ be defined by
\begin{equation}
    \mathcal{S}(\mathcal{B}_{L}) = \Gamma\setminus\bigcup_{b\in\mathcal{B}_{L}}b,
    \label{singletons}
\end{equation}
with $\setminus$ being the set minus operator. \\For $b=(k_1,\ldots,k_L)\subseteq \Gamma$, $k_i\in \Gamma$, $i=1,\ldots,L$, let $\boldsymbol{\theta}_{b} = (\theta_{k_1},\ldots,\theta_{k_L})^{\top}$. A Morph approximation of order $L$ is a product approximation of $P(\boldsymbol{\theta})$ such that 
\begin{equation}
    \mathcal{M}_{\mathcal{B}_{L}}(\boldsymbol{\theta})=\prod_{b\in\mathcal{B}_{L}}P_{b}(\boldsymbol{\theta}_{b})\prod_{s\in \mathcal{S}(\mathcal{B}_{L})}P_{s}(\theta_{s}),
    \label{eq:M_b}
\end{equation}
with $P_{b}$ the joint probability distribution of the sub-vector~$\boldsymbol{\theta}_{b}$ of length $L$.  
\end{definition}
Although all $\mathcal{M}_{\mathcal{B}_{L}}(\boldsymbol{\theta})$ are valid product approximations of $P(\boldsymbol{\theta})$, these approximations differ in their closeness to $P(\boldsymbol{\theta})$. Therefore, one can choose an optimal Morph approximation where $\mathcal{M}_{\mathcal{B}_{L}}$ covers the important dependencies between variables.
Accordingly, the total correlation, also known as multi-information \cite{watanabe1960information, Studený1998}, is a suitable measure to weight the importance of the dependencies between variables in each $\boldsymbol{\theta}_{b},\,\forall b\subseteq \Gamma$. The total correlation $\mathcal{C}$ of $\boldsymbol{\theta}_{b}$ as a function of entropy $H$ is given by 
\begin{equation}
    \mathcal{C}(\boldsymbol{\theta}_{b}) = D_{KL}\left(P_{b}||\prod_{k\in b}P_{k}\right) = \sum_{k\in b}H(\theta_{k})-H(\boldsymbol{\theta}_{b}),    
\end{equation}
where $D_{KL}$ is the \textit{Kullback-Leibler divergence} between $P_{b}$, the joint probability density, and $\prod_{k\in b}P_{k}$, the element-wise product of the marginal densities of the block $b$. The \textit{Kullback-Leibler divergence} between $P(\boldsymbol{\theta})$ and $\mathcal{M}_{\mathcal{B}_{L}}(\boldsymbol{\theta})$ as a function of $\mathcal{C}$ can be represented by
\begin{equation}
     D_{KL}\left(P||\mathcal{M}_{\mathcal{B}_{L}}\right) = \mathcal{C}(\boldsymbol{\theta})-\sum_{b\in B}\mathcal{C}(\boldsymbol{\theta}_{b}),  
\end{equation}
where $\mathcal{C}(\boldsymbol{\theta})$ is the total correlation of the parameter vector $\boldsymbol{\theta}$ and $\sum_{b\in B}\mathcal{C}(\boldsymbol{\theta}_{b})$ the sum of the total correlations of  the disjoint-blocks (see appendix~\ref{app:kL TC derivation} for the derivation).
Therefore, one can use $\sum_{b\in B}\mathcal{C}(\boldsymbol{\theta}_{b})$ as a measure of the total covered dependencies in $\mathcal{M}_{\mathcal{B}_{L}}(\boldsymbol{\theta})$ where its maximization increases the closeness of $\mathcal{M}_{\mathcal{B}_{L}}(\boldsymbol{\theta})$ and $P(\boldsymbol{\theta})$.
\begin{definition}\label{def:optimal l morph approx.}
$\mathcal{M}_{\mathcal{B}^{*}_{L}}$ is an optimal Morph approximation of order $L$ if
\begin{equation}
    \mathcal{B}^{*}_{L} \in \text{argmax}_{\mathcal{B}_{L}\in\mathcal{F}_{L}}\sum_{b\in \mathcal{B}_{L}}\mathcal{C}(\boldsymbol{\theta}_{b}).
\end{equation}

\end{definition}

An optimal Morph distribution of order $L$ is, by definition, normalized and incorporates the maximum dependencies between variables in blocks of length $L$, while maintaining a simple approach.

\subsection{\morphz}
\label{subsec:morphz}

We introduce \morphz, an estimator for marginal likelihood, which uses the Morph approximation as a proposal distribution for the bridge sampling (BS) estimator \cite{meng1996simulating}. 
The bridge sampling estimator is given by 
\begin{equation}
    \textit{z}=\frac{\mathbb{E}_{g(\boldsymbol{\theta})}[L(X|\boldsymbol\theta)\pi(\boldsymbol\theta)h(\boldsymbol{\theta})]}{\mathbb{E}_{post}[h(\boldsymbol{\theta})g(\boldsymbol{\theta})]},
    \label{eq:BS}
\end{equation}
where $h(\boldsymbol{\theta})$ is the bridge function, $g(\boldsymbol{\theta})$ is the proposal distribution (or the importance distribution when $h(\boldsymbol{\theta})=1$) and $\mathbb{E}_{g(\boldsymbol{\theta})}$ and $\mathbb{E}_{post}$ are the expected values with respect to the proposal and posterior distribution, respectively.
Specifically, the bridge function is set to the optimal bridge function \cite{meng1996simulating} defined by 
\begin{equation}
    h(\boldsymbol{\theta})= C\cdot\frac{1}{f_{1}\;L(X|\boldsymbol\theta)\pi(\boldsymbol\theta)+f_{2}\;\textit{z}\; g(\boldsymbol{\theta})},
    \label{eq:BS_bridge}
\end{equation} 
with $f_{1}=N_1/(N_1+N_2)$ and $\text{f}_{2}=N_2/(N_1+N_2)$ the fraction of $N_1$ posterior and $N_2$ proposal samples, respectively, and $C$ a constant that cancels out in the ratio in Equation~\ref{eq:BS}. An iterative scheme is used to estimate the marginal likelihood from an initial guess of $\textit{z}$ \cite{gronau2017tutorial}. The accuracy of BS estimates can be quantified through an approximation of the relative mean-squared error~\cite{fruhwirth2004estimating}, where both posterior and proposal samples can be used at no-extra sampling cost~\cite{gronau2017tutorial}.  A Python-based implementation~\footnote{\morphz: \url{https://github.com/EL-MZ/MorphZ}} of the BS estimator has been developed for this study.  Our implementation follows the detailed description of the bridge sampling estimator and its relative-error (RE) in \citeauthor{gronau2017tutorial} \cite{gronau2017tutorial}. We use the Morph approximation $\mathcal{M}_{\mathcal{B}_{L}}(\boldsymbol{\theta})$ as a proposal distribution $g(\boldsymbol{\theta})$. For simplicity, we will refer to $\mathcal{M}_{\mathcal{B}_{L}}(\boldsymbol{\theta})$ as $\mathcal{M}_{L}(\boldsymbol{\theta})$. We adopt the sample-splitting technique~\cite{overstall2010default}, which consists of dividing the available posterior samples into one batch for constructing the Morph approximation and another for the bridge sampling phase. 

The kernel density estimator (KDE) is a fast standard method for density estimation of lower-dimensional problems~\cite{sheather1991reliable,botev_2010}.
The Morph approximation factorizes a higher-dimensional probability density into a product of low-dimensional factors, each of which can be estimated efficiently with KDEs. Throughout this work, the multivariate KDE with a Gaussian kernel~\cite{virtanen2020scipy} is used to estimate the densities $P_{b}(\boldsymbol{\theta}_{b})$ and $P_{s}(\boldsymbol{\theta}_{s})$ which are used both in $\mathcal{M}_{L}(\boldsymbol{\theta})$ and $\mathcal{C}(\boldsymbol{\theta}_{b})$.

\begin{algorithm}[t]
\small
\caption{\textsc{Seeded Greedy Maximization}}
\label{sgm}
\textit{Select $m$ pairwise-disjoint blocks of length $L$ maximizing $\sum_{b\in \mathcal{B}_{L}} C(\boldsymbol{\theta}_{b})$.}

\medskip
\textbf{Inputs:} Dimension $n$; block length $L$; number of seeds $K\ge 1$; total correlation $C(\boldsymbol{\theta}_{b})\ge 0$ for each block $b$. 
Let $\mathcal{W}$ be the full collection of $b$ candidates, i.e., \emph{all $\binom{n}{L}$ length $L$ combinations of the $n$ dimensions}.\\

\textbf{Output:} A set $B_{\mathrm{best}}\subseteq\mathcal{B}_{L}$ of at most
$m$ pairwise-disjoint blocks with the largest $\sum_{b\in \mathcal{B}_{L}} C(\boldsymbol{\theta}_{b})$ between the top $K$ seeds.

\medskip
\begin{enumerate}[leftmargin=*,label=\arabic*.]
\item \textbf{Precompute and sort.}
  Set $m\leftarrow \lfloor n/L\rfloor$.
  Let $\mathcal{W}\leftarrow$ the list of all possible blocks $b$ sorted in decreasing $C(\boldsymbol{\theta}_{b})$.
  Define the predicate:   $\textsf{disjoint}(b,\textit{used})\iff b\cap \textit{used}=\varnothing
  $.

\item \textbf{Loop over the top $K$ seeds.}
  For $j=1,\dots,\min\{K,|\mathcal{W}|\}$:
  \begin{enumerate}[leftmargin=*,label=(\roman*),nosep]
    \item \emph{Start a fresh construction.}\\
      \quad $B\leftarrow \varnothing$;\quad $\textit{used}\leftarrow \varnothing$;\quad $\textit{total}\leftarrow 0$.
    \item \emph{Place the seed.}\\
      $\textit{seed}\leftarrow \mathcal{W}[j]$; $B\leftarrow B\cup\{\textit{seed}\}$;\;
      $\textit{used}\leftarrow \textit{used}\cup \textit{seed}$;\;
      $\textit{total}\leftarrow \textit{total}+C(\boldsymbol{\theta}_{\textit{seed}})$.
    \item \emph{Greedy fill.}\\
      Scan $\mathcal{W}\setminus\{\textit{seed}\}$ in order while $|B|<m$.\\
      For each $b$:
      if $\textsf{disjoint}(b,\textit{used})$ then
      $B\leftarrow B\cup\{b\}$;\;
      $\textit{used}\leftarrow \textit{used}\cup b$;\;
      $\textit{total}\leftarrow \textit{total}+C(\boldsymbol{\theta}_{b})$.
    \item \emph{Keep the best construction.}\\
      If $\textit{total}>\textit{Best total}$ then
      $\textit{Best total}\leftarrow \textit{total}$ and
      $B_{\mathrm{best}}\leftarrow B$.
  \end{enumerate}

\item \textbf{Return} $B_{\mathrm{best}}$.
\end{enumerate}

\medskip

\end{algorithm}

To construct $\mathcal{M}_L(\boldsymbol{\theta})$, we compute $\mathcal{C}(\boldsymbol{\theta}_b)$ for
every candidate block $b\subset\Gamma$.  Then, a seeded greedy maximization, SGM (see Algorithm~\ref{sgm}): sort all blocks by
\emph{decreasing} $\mathcal{C}$; iterate over the top $K$ blocks as seeds; for each seed, greedily add the highest-scoring non-overlapping blocks until $m$ blocks are selected; return the run with the largest \deleted{total score} \added{sum of total correlations}. \deleted{The SGM achieves a $1/L$-approximation near optimal solution to the maximum-weight selection, thus avoiding the NP-hard exact search for $L>2$ while remaining fast and scalable \mbox{\cite{2006mestre}.}}
\added{For $L=2$, where the optimum is computationally tractable, SGM achieves or closely approaches optimality on our benchmarks; while for $L>2$, it maintains similarly high performance relative to the best computed sum of total correlations (see \mbox{Figure~\ref{fig:sgm_ilp_accuracy_runtime})}.}
The product structure of the Morph approximation allows each factor to propose samples independently. Taking one draw from each factor gives joint samples that one can evaluate under the posterior. In practice, fast KDEs enable us to generate large arrays of samples almost instantly. We leverage this feature to vectorize KDE draws and assemble batched joint proposals in a single pass. 

To compare the efficiency of the Morph approximation under different sampling methods, we introduce a particular case of the GSS method \cite{fan2011choosing}  that incorporates the Morph approximation as the proposal distribution. We refer to this method as \textit{Morphed Steppingstone Sampling} (MSS). The Python implementation of GSS presented in Ref.~\cite{zahraoui2025generalized}, combined with the Morph approximation, is utilized in Section~\ref{appli:stats}. 

\begin{table*}
    \centering
    \setlength{\tabcolsep}{7.5pt} 
    \begin{tabular}{cccccccccccc}\toprule
         \multicolumn{3}{c}{}&  \multicolumn{3}{c}{Nested Sampling}&  \multicolumn{3}{c}{\morphz}&  \multicolumn{3}{c}{MSS}\\\midrule
         Model&$d$& $\log(z)_{\text{True}}$&  calls&  $|\Delta|$&  std&  calls&  $|\Delta|$&  std&  calls&  $|\Delta|$&std\\
         \textit{Egg-box}&   2&235.856&  $6.9\times10^{4}$&  0.025&  0.106& $5\times10^{3}$&  0.035&  0.01&  $4\times10^{3}$&  0.062&0.72\\
 \textit{Peak-plateau}&  20&0.693& $1.8\times10^{6}$& 0.332& 0.277& $5\times10^{3}$& 0.051& 0.06& $4\times10^{3}$& 1.555&1.6\\
 \textit{Gaussian-shells}&  30&-60.13& $6.5\times10^{6}$& 0.921& 0.317& $5\times10^{3}$& 0.074& 0.01& $4\times10^{3}$& 0.141& 0.35\\ \bottomrule
    \end{tabular}
    \caption{A comparison of the accuracy as $|\Delta|=|\overline{\log(\hat{z})}_{\text{Method}}- \log(z)_{\text{True}}|$, along the standard deviation of 100 $\log(\hat{z})$, and the efficiency as the number of likelihood calls for each estimate between Nested Sampling, \morphz, and MSS for the 3 statistical examples.
    }
    \label{tab:benchmarks}
\end{table*}

\section{Application to statistical examples}
\label{appli:stats}

In this section, we compare both \morphz and MSS to nested sampling through challenging statistical problems that reflect real inference scenarios. These statistical problems are considered benchmarks for assessing the accuracy of marginal likelihood estimation. Two of these benchmarks consist of the egg-box likelihood and the Gaussian shells likelihood~\cite{feroz2013importance}. As an additional benchmark, a different version of the peak likelihood originally discussed by~\citeauthor{skilling2006nested}\mbox{\added{~\citep{skilling2006nested}}}, represents one of the most challenging problems for evidence estimation methods. In this case, only Nested sampling can efficiently draw samples from the posterior distribution, yet requires a higher cost for an accurate marginal likelihood estimate~\cite{russel2019model}.  

We set up each model, where the data is a null vector, as follows:
\begin{itemize}
    \item \textbf{\textit{Egg-box}}: The two-dimensional likelihood is given by 
    \begin{equation}
        L(\boldsymbol{\theta})= \text{exp}\left[\left(2+\text{cos}\left(\frac{\theta_1}{2}\right)+\text{cos}\left(\frac{\theta_2}{2}\right)\right)^{5}\right],
        \label{ln_like:eggbox}
    \end{equation}
    with uniform priors~$\mathcal{U}(0,10\pi)$ for both parameters.
    
    \item \textbf{\textit{Gaussian-shells}}: The $d$-dimensional likelihood is defined as
    \begin{equation}
        L(\boldsymbol{\theta})= \text{circ}(\boldsymbol{\theta},\boldsymbol{c_1},r_1,w_1)+\text{circ}(\boldsymbol{\theta},\boldsymbol{c_2},r_2,w_2),
        \label{ln_like:shell}
    \end{equation}
    where the $\text{circ}$ function is given by
    \begin{equation}
        \text{circ}(\boldsymbol{\theta},\boldsymbol{c},r,w)=\frac{1}{\sqrt{2\pi w^{2}}}\text{exp}\left[-\frac{(|\boldsymbol{\theta}-\boldsymbol{c}|-r)^{2}}{2w^{2}}\right],
        \label{ln_like:circ}
    \end{equation}
    with $\boldsymbol{\theta}$ the parameter vector, $\boldsymbol{c}$ the center point, $r$ the radius, and $w$ the Gaussian radial profile of width.
    Following~\cite{feroz2009multinest}, we set the model to $d = 30 $, $r_1=r_2= 2$, and $w_1=w_2=0.1$ with a uniform prior~$\mathcal{U}(-6,6)^{d}$.
    \item \textbf{\textit{Peak-plateau}}: The $d$-dimensional likelihood is expressed as
    \begin{align}
        L(\boldsymbol{\theta})=\prod_{i=1}^{d}\frac{1}{v\sqrt{2\pi}}\text{exp}\left[-\frac{\theta_{i}}{2v^{2}}\right]+\\ \nonumber
        A\prod_{i=1}^{d}\frac{1}{u\sqrt{2\pi}}\text{exp}\left[-\frac{\theta_{i}}{2u^{2}}\right],
        \label{ln_like:pp}
    \end{align}
    where $\boldsymbol{\theta}$ is the parameter vector, $v=0.1$ the variance for the Gaussian plateau, $u=0.01$ the variance for the Gaussian peak, and $A=1$ the scale of the peak with uniform priors~$\mathcal{U}(-0.05,0.05)^{d}$\added{~\cite{russel2017bayesian}} .
\end{itemize}

\begin{figure}
    \centering
    \includegraphics[width=1\linewidth]{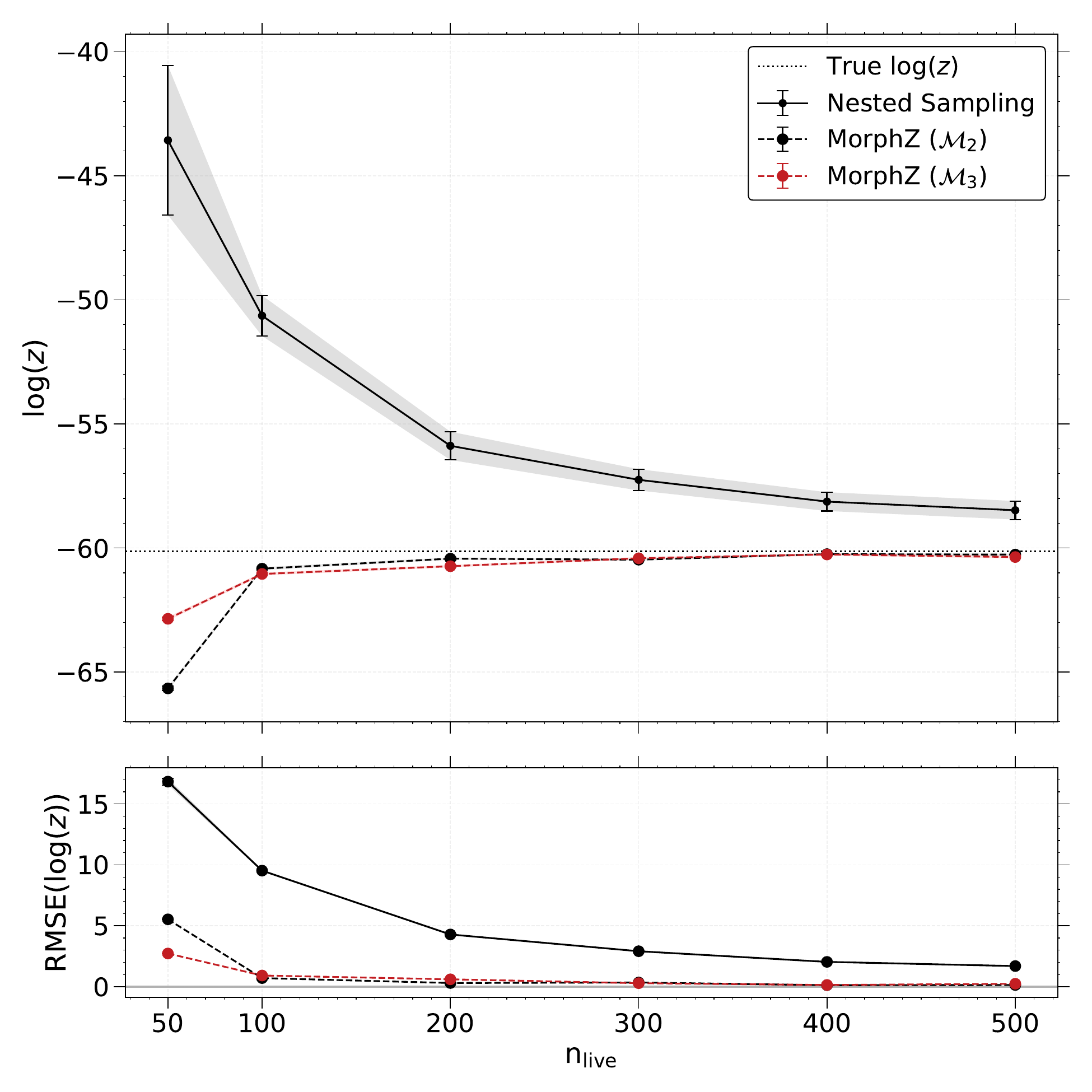}
    \caption{Convergence and accuracy of \textit{Gaussian-shells} $\log(z)$ estimates versus $n_{\rm live}$ the number of live points. The posterior samples from one NS run with $n_{\rm live}$ is used for \morphz estimates plotted at  $n_{\rm live}$, with 3000 likelihood calls (for BS proposal samples) per each \morphz estimate.
\textit{Top:} Estimated $\log(z)$, NS (black) and the \morphz with Morph approximations of second order~($\mathcal{M}_2$, gray) and third order~($\mathcal{M}_3$, red). 
Symbols denote means over 100 run; error bars indicate $1\sigma$ uncertainties. 
The horizontal dotted line marks the true $\log(z)$.
\textit{Bottom:} Root-mean-square error (RMSE) of $\log(z)$ relative to the true value .
        }
    \label{fig:morph vs nlive}
\end{figure}
For all models, the nested sampling (NS) estimates are obtained using \dynesty \cite{skilling2006nested,speagle2020dynesty,koposov2023joshspeagle}.  For each model, we run NS with the number of live points set to $n_
{\rm live}=500$ and an adequate stopping criterion. Since the Morph approximation requires posterior samples, we reuse the posterior samples produced from one of the NS runs for each model. For the \morphz, we construct a second-order Morph approximation $\mathcal{M}_{2}$, with the standard \textit{Silverman}'s rule for the KDE bandwidth. A total of 4000 NS posterior samples are split evenly: one half to construct $\mathcal{M}_{2}$, and the other half for the bridge sampling. For the MSS, we reuse the same Morph approximation and set the number of temperatures to $4$. 
\begin{table*}
    \centering
    \setlength{\tabcolsep}{7pt} 
    \begin{tabular}{ccccccccc}\toprule
          \multicolumn{3}{c}{}&   \multicolumn{3}{c}{GSS}& \multicolumn{3}{c}{\morphz}\\\midrule
          PTA&Model Name&  $d$&    calls&$\overline{\log(\hat{z})}$&std& calls&$\overline{\log(\hat{z})}$&std  \\
          NANOGrav&PSR J1705-1903 WN+RN&  8&    $8\times10^{3}$&111 562.63&0.44& $1\times10^{1}$&111 562.60& 0.06\\
          NANOGrav&PSR J1012+5307 WN+RN+DMGP&  16&    $8\times10^{3}$&275 745.76&0.61& $2\times10^{1}$&275 745.53& 0.06\\
 NANOGrav& PSR J1713+0747 WN+RN& 32&  $8\times10^{3}$&751 285.52& 0.78& $3\times10^{2}$&751 284.62& 0.05\\
 EPTA& DR2 NEW PSRN+GWB& 70& $3.2\times10^{4}$&  493 692.55& 1.14& $5\times10^{2}$& 493 693.40& 0.39\\
          EPTA&DR2 FULL PSRN+CURN&  74&    $3.2\times10^{4}$&607 134.56&1.20& $5\times10^{2}$&607 135.48& 0.15\\
 NANOGrav& 15 yr CURN& 136&  $8\times10^{4}$&7 353 042.72& 0.84& $3\times10^{3}$&7 353 041.33& 0.32\\
 NANOGrav& 15 yr HD& 136& $8\times10^{4}$& 7 353 047.80&  0.91& $4\times10^{3}$& 7 353 046.79&0.31\\ \bottomrule
    \end{tabular}
    \caption{Comparison of $\log(z)$ estimates using GSS, \morphz for different PTA models with dimension $d$. The mean and standard deviation of 100 $\log(z)$ estimates per each model are displayed along the number of likelihood calls per estimate.}
    \label{tab:pta}
\end{table*}

\deleted{\mbox{Table~\ref{tab:benchmarks}} reports the mean and standard deviation of 100 $\log(z)$ estimates for each method on the three benchmarks, with the number of likelihood evaluations (calls) used for each single estimate.}
\added{\mbox{Table~\ref{tab:benchmarks}} reports the true evidence $\log(z)_{\text{True}}$, the absolute distance between the mean $\overline{\log{\hat{z}}}$ of 100 estimate and $\log(z)_{\text{True}}$, the standard deviation of the estimates, and the number of likelihood calls (evaluations) used for each single estimate.}
\
For the \textit{Egg-box} problem, both Morph-based methods  produce $\log(z)$  estimates consistent with the NS estimate, demonstrating robustness to the high multi-modality of this likelihood. For the \textit{Gaussian-shells} problem, \morphz and MSS outperform NS. Their evidence estimates are more accurate and obtained at roughly two orders of magnitude lower computational cost. For the \textit{Peak-plateau}, only the \morphz estimator succeeded in producing an estimate closer to the true $\log(z)$, while NS requires a higher number of live points $n_
{\rm live}$ to achieve similar accuracy.

Since MSS is based on path sampling, the \textit{Peak-plateau} is sharply peaked over a narrow region but nearly flat over a wide plateau, so accurate path sampling requires a substantially \deleted{large} \added{larger} number of likelihood evaluations to explore both regions adequately~\cite{skilling2006nested}. \citeauthor{russel2017bayesian} \citeyear{russel2017bayesian} \cite{russel2017bayesian} shows that GSS can accurately estimate the marginal likelihood of the \textit{Peak-plateau} problem with $A=100$ using 48 temperatures. However, for $A=1$ GSS fails to estimate $\log(z)$ even with 100 temperatures. In this case, the tails contain more probability than the spike region compared to the case of  $A=100$. In the current setup, MSS underestimates $\log(z)$ with high variance. Across all models, we have shown that \morphz estimated the marginal likelihood accurately and at a lower cost compared to nested sampling and MSS, while MSS has a slightly higher variance than \morphz. Moreover, MSS requires the likelihood, as a function of the prior mass, to be concave to avoid failure. 
In practice, the likelihood as a function of the prior mass is usually approximately concave, in which case path sampling is effective.

In Figure~\ref{fig:morph vs nlive}, we compare nested sampling  and \morphz estimates for the \textit{Gaussian-shells} example. First, NS is used to obtain posterior samples with increasing $\text{n}_{live}$ the number of live points. Then at each $\text{n}_{live}$, \morphz uses the same posterior samples drawn from one of the NS runs to construct second and third order Morph approximation (respectively $\mathcal{M}_2$ and $\mathcal{M}_3$ to show \morphz consistency), along with 3000 \deleted{additional} likelihood \deleted{evaluations} \added{calls} \added{(for BS proposal samples)}  per $\log(z)$ estimate. As $\text{n}_{live}$ increases, the NS estimates converge gradually to the true $\log(z)$ reaching a \deleted{$|\Delta\log(z)|< 1.7$} \added{$|\Delta|< 1.7$} at $\text{n}_{live}=500$. In contrast, both \morphz approximations $\mathcal{M}_2$ and $\mathcal{M}_3$ are already consistent with the true value with \deleted{$|\Delta\log(z)|< 0.9$} \added{$|\Delta|< 0.9$} at $\text{n}_{live}=100$ and remain more accurate thereafter. The RMSE shows that NS errors decrease with $\text{n}_{live}$ yet remain significantly higher than those of \morphz. Note that even at $\text{n}_{live}=50$, where the posterior coverage is incomplete, \morphz complements NS by leveraging the available posterior information provided by NS and closing the gap to an accurate estimate. 

One can use NS with a reduced number of live points to explore parameter space and generate posterior samples, and then apply \morphz in post-processing to obtain an estimate closer to the true $\log(z)$, substantially cutting computational costs.

\section{Application to the Pulsar Timing Array}
\label{appli:pta}
Recently, GSS was applied to different PTA analyses, demonstrating substantial reduction in computational cost for accurate evidence estimation \cite{zahraoui2025generalized}. In this application, we compare the efficiency of \morphz for model evidence estimation across problems whose dimensionality ranges from $8$ to $136$. A second-order Morph approximation $\mathcal{M}_{2}$ is constructed using SGM, based on 2000 posterior samples for $8-32$ dimensions and 5000 samples for $70-136$ dimensions, and adopt \textit{Silverman}'s rule for the KDE bandwidth. In these settings, SGM iterates over the complete set of pairwise seeds. For \morphz, the bridge-sampling settings are those discussed in Section~\ref{subsec:morphz}.
\begin{figure}
    \centering
    \includegraphics[width=1\linewidth]{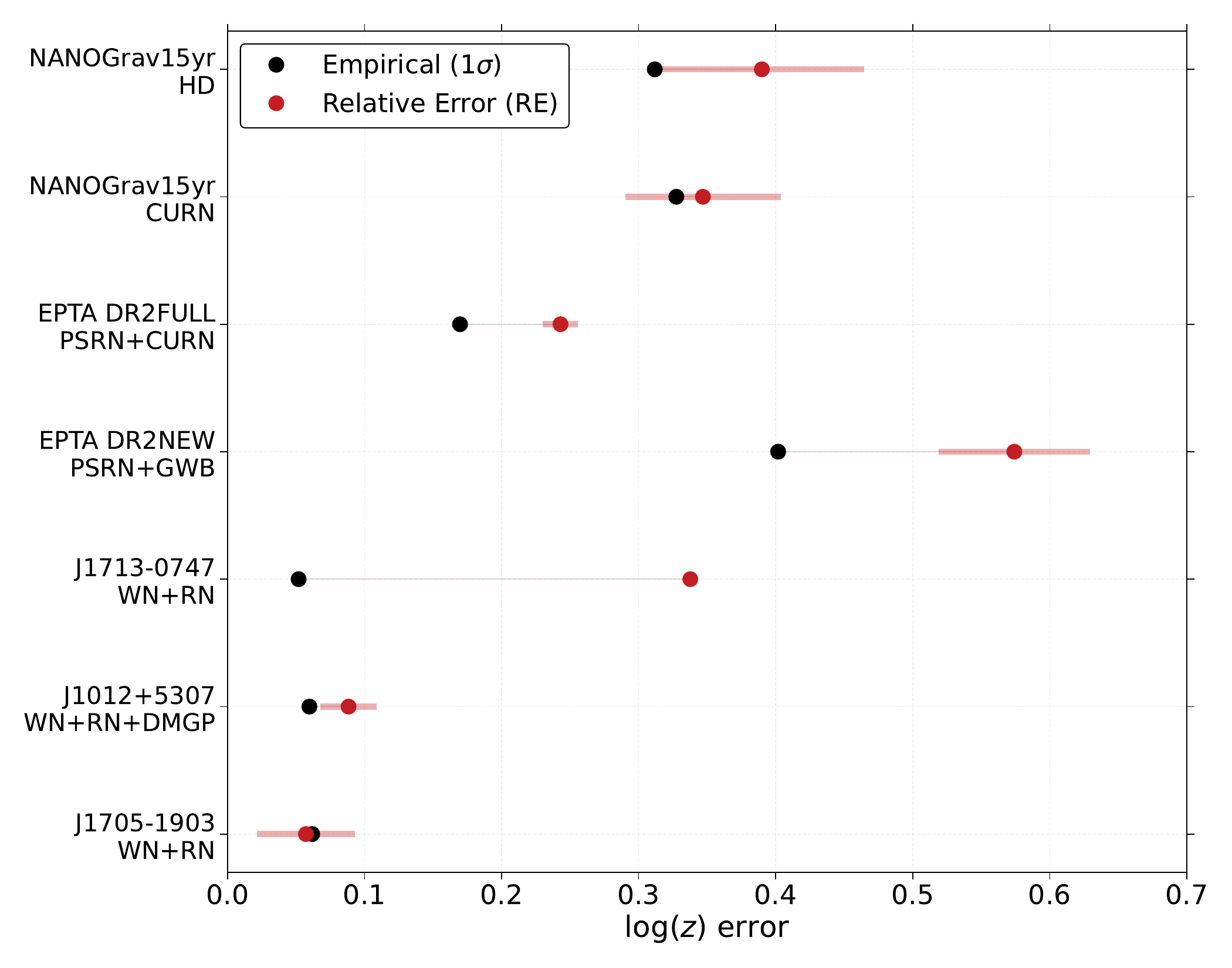}
    \caption[Empirical vs. BS RE $\log(z)$ error across models]{Comparison of bridge sampling RE and empirical errors for $\log(z)$ across models in Table~\ref{tab:pta}. 
For each model, the red dot marks and the faint red bar shows respectively the mean and $\pm 1 \sigma$ of 100 RE, and the black square marks the empirical error as the standard deviation of $100$ estimates of $\log(z)$. 
The horizontal red segment connects the approximate and empirical values; indicating the degree of under/overestimation.}
\label{fig:logz-error-calibration}
\end{figure}
\begin{figure}
    \centering
    \includegraphics[width=1\linewidth]{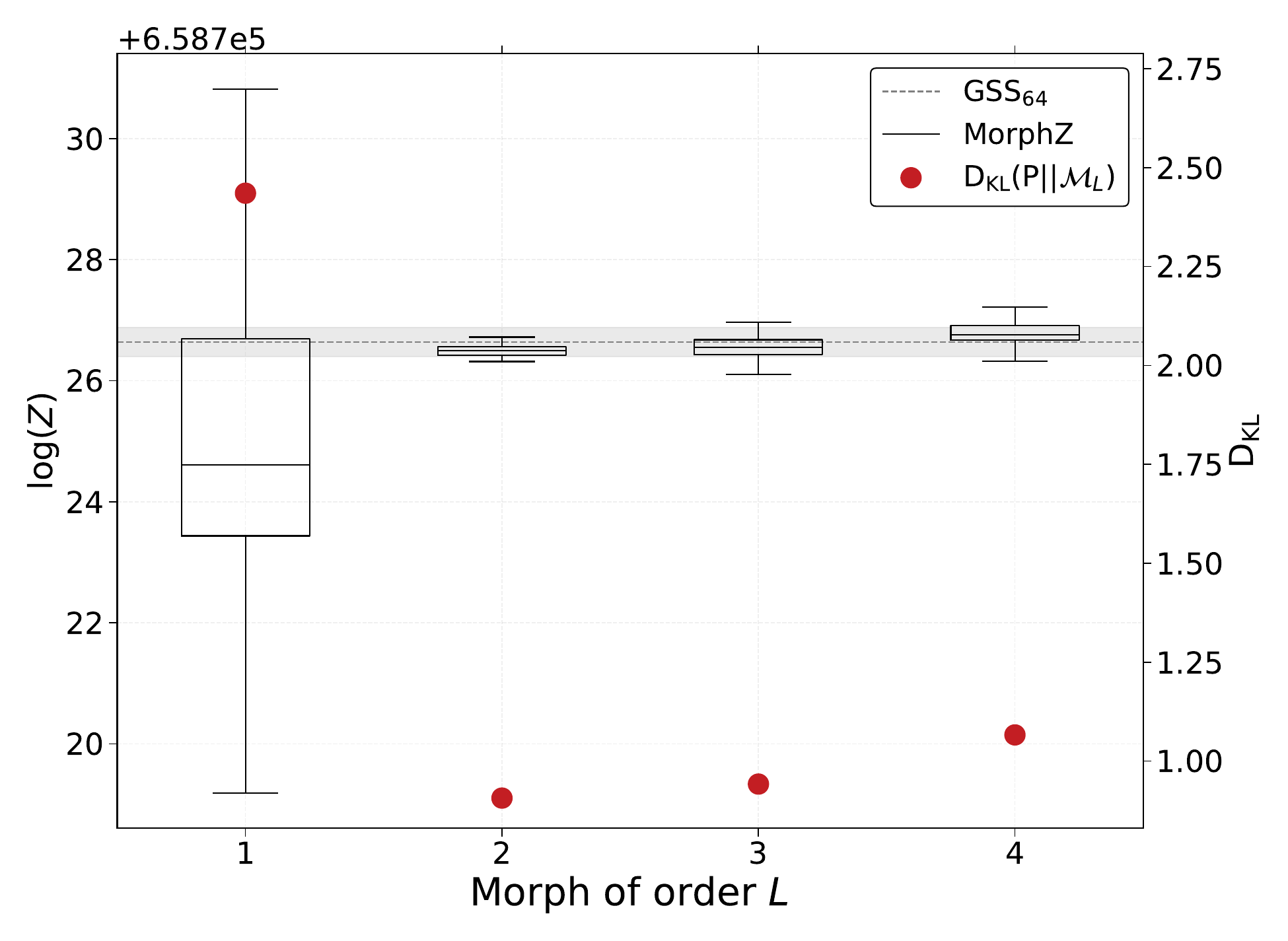}
    \caption{Comparison of $\log(z)$ estimates for the EPTA DR2FULL PSR+CURN model using GSS and \morphz. The box plots show the distribution of $\log(z)$ for each Morph approximation order. The horizontal dashed line and shaded band indicate the mean and $\pm1\sigma$ range of the reference GSS estimates with 64 temperatures. Red diamonds (right axis) denote the forward Kullback-Leibler divergence $D_{\mathrm{KL}}(P|\mathcal{M}_{\mathcal{B}_{L}})$ between the posterior and each corresponding order of the Morph approximation.}
    \label{fig:morph_Dkl}
\end{figure}

Table~\ref{tab:pta} presents the mean and standard deviation of 100 $\log(z)$ estimates computed with GSS and \morphz, together with the number of likelihood evaluations per estimate ($\text{calls}$). For low-dimensional models $d=8-16$, \morphz produced an accurate $\log(z)$ estimate using only 10-20 \deleted{additional} likelihood evaluations. For mid-dimensional models $d=32-74$, \morphz consistently estimated $\log(z)$ with $300-500$ samples, even for costly models like the  \texttt{EPTA DR2 NEW PSRN+GWB} model. For high-dimensional problems $d=136$, \morphz reduced the computational cost by a factor of $ 20$ compared to GSS, making it particularly attractive for very expensive models such as the \texttt{NANOGrav 15yr HD} model. 

In Figure~\ref{fig:logz-error-calibration}, we compare the empirical error of $\log(z)$, defined as $1\sigma$ of 100 estimates, with the BS relative-error (RE) of each of the 100 estimates. Across all the models listed in Table~\ref{tab:pta}, the relative error is either consistent, or is slightly overestimated compared to the empirical one. 
This mild overestimation can act as a safeguard, making RE conservative when empirical error is expensive to obtain, as in the case of \texttt{NANOGrav 15yr HD}. Overall, \morphz achieves a two to three orders of magnitude reduction in likelihood calls relative to GSS, while providing more accurate $\log(z)$ and conservative relative error estimates, demonstrating efficiency over problems with varying difficulties.

In Figure~\ref{fig:morph_Dkl}, we compare different orders of  the Morph approximation used in \morphz to estimate the marginal likelihood of  \texttt{EPTA DR2 NEW PSRN+CURN} model. A reference $\log(z)$ is estimated using GSS with 64 temperatures and $30,000$ samples per temperature. For second and higher orders of the Morph approximation, the $\log(z)$ estimates are consistent with those from GSS while exhibiting similar variance at a cost of only 3000 likelihood evaluations \ per estimate. The Kullback-Leibler divergence shows that incorporating dependencies between posterior samples lowers the distance between the posterior and $\mathcal{M}_{L\geq2}$. This additional information~leads to more accurate estimates of $\log(z)$ and a reduction in its variance. The comparison of $\text{D}_{\text{KL}}(P(\boldsymbol{\theta})||\mathcal{M}_{L\geq2}(\boldsymbol{\theta}))$ does not, however, accurately reflect the theoretical prediction, because an exactly optimal Morph approximation can only be obtained for $L = 2$. We discuss this in more detail in Section~\ref{discuss}.

\section{Application to LVK Compact Binary Coalescences}
\label{appli:LVK}


\begin{table*}[t]
\centering
\caption{Ensemble sumary of $\Delta \log(\hat{z})$ per method relative to \lnZns \ over 100 BBH simulations (see Figure~\ref{fig:logz-error-lvk}). IQR denotes the 25--75\% interval. }
\label{tab:ensemble-deltas}

\begin{tabular}{ccccccc}
\toprule
Method & Median &IQR$_{25-75\%}$& Mean&std& $|\Delta\log(\hat{z})|<0.5$ & $|\Delta\log(\hat{z})|<1.0$ \\ 
\midrule
SS & 1.049&0.316\,-\,1.629& -0.373&4.788& 12.1\% & 30.3\% \\
MorphZ$_{(\mathrm{NS})}$ & 0.034&-0.130\,-\,0.153& 0.010&0.226& 96.0\% & 100.0\% \\
MorphZ$_{(\mathrm{PT})}$ & 0.096&-0.005\,-\,0.223& 0.106&0.219& 94.9\% & 100.0\% \\
\bottomrule
\end{tabular}
\end{table*}

In this section, we assess the accuracy of \morphz for compact binary coalescence (CBC) and the GW150914 event for gravitational-wave inference. We compare \morphz evidence estimates against nested sampling (NS;~\cite{skilling2006nested,speagle2020dynesty,koposov2023joshspeagle}) and Steppingstone Sampling (SS; \cite{maturana2019stepping}) results while maintaining identical data, priors, likelihoods, and waveform models. All CBC inferences are carried out using \bilby~\cite{Ashton2019,RomeroShaw2020}. To ensure full reproducibility, all scripts, prior definitions, and detailed sampler configuration used in this study are provided in the accompanying code repository (see Section~\ref{data}).

We consider two datasets.  
(i) A simulation study of 100 independent binary black hole (BBH) signal injections, each corresponding to a distinct synthetic GW source with different intrinsic parameters. The extrinsic parameters ($\psi$, RA, Dec, coalescence time, phase, luminosity distance, and inclination) are fixed to provide a controlled comparison with reduced extrinsic uncertainty.  
(ii) GW150914, analyzed with distance and time marginalization. 
The Power Spectral Density (PSD) for GW150914 are obtained from the public LVK parameter-estimation release~\cite{GWTC2data, GWTC2paper}, while the simulation study uses the \bilby default design-sensitivity PSD.
\begin{figure}
    \centering
    \includegraphics[width=0.95\linewidth]{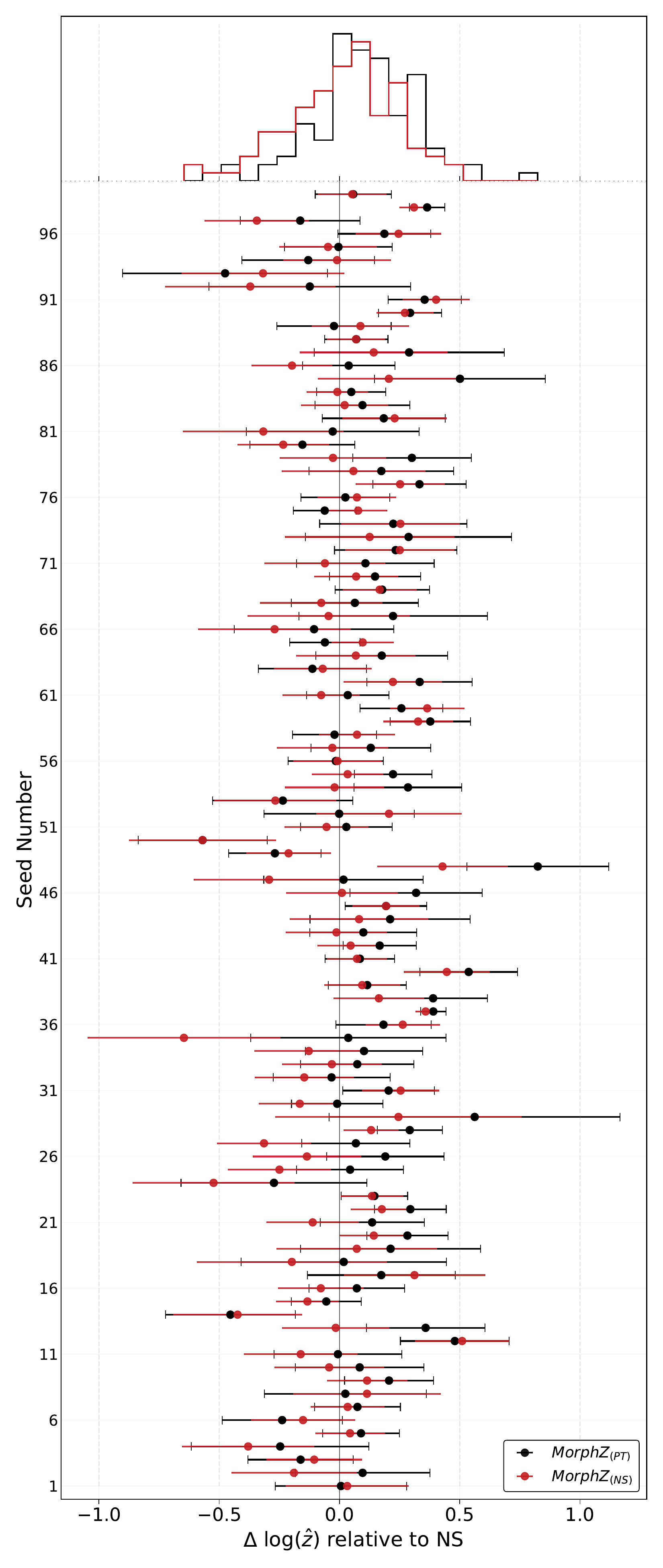}
    \caption{Comparison of \dynesty, \morphz$_{\text{(PT)}}$, and \morphz$_{\text{(NS)}}$ across 100 independent BBH simulation (see Table~\ref{tab:ensemble-deltas}). The mean and standard deviation of 100 $\log(z)$ estimate per simulation is used to compute  $\Delta \log(\hat{z})= \log(\hat{z})_{\mathrm{method}} - \log(\hat{z})_{\mathrm{NS}}$ for both independent \morphz estimates. Note that $\log(z)_{\mathrm{NS}}$ is used as a reference estimate because the true $\log(z)$ is unknown.}
\label{fig:logz-error-lvk}
\end{figure}
 The 100 simulated CBC analyses uses 4\,s of data ($ injection(\text{CBC GW}) + noise(\text{Design-sensitivity PSD}) $), and the standard frequency-domain CBC Whittle likelihood with distance and time marginalization enabled~\cite{ThraneTalbot2019IntroToBayes}. 
We employ the \texttt{IMRPhenomPv2} waveform model~\cite{Sascha2016IMRPhenomPv2,Khan2016IMRPhenomPv2} and analyse data from the LIGO Hanford (H1) and Livingston (L1) detectors~\cite{Abbott2016_AdvLIGO}. 
Nested sampling is performed with \dynesty using $n_{\mathrm{live}} = 2000$, $n_{\mathrm{act}} = 20$, and the \bilby implementation of the \texttt{rwalk} proposal. Additionally, parallel-tempered (PT) \bilbymcmc~\cite{Ashton2021} is used independently with $n_\mathrm{temps}=8$, to retain posterior samples with $n_\mathrm{samples}=2000$ after adaptive tuning, thinning by $\gamma=0.2$, proposal cycle \textsc{gwA} \added{as implemented in \bilbymcmc~\cite{Ashton2021}}, and $(L1,L2)=(100,5)$ sub-steps. The temperature ladder is set using the \texttt{Tmax\_from\_SNR} prescription with $T_{\max}=20$. 
SS is employed to estimate the evidence using the \bilbymcmc runs, a setting known to retain residual bias unless a finer temperature resolution is used.

\begin{table*}[t]
\setlength{\tabcolsep}{7pt} 
\centering
\caption{ $\Delta\log(\hat{z})$ summary per method  relative to \lnZns\;for the GW150914 event.}
\label{tab:gw1509}

\begin{tabular}{ccccccccc}\toprule
 &\multicolumn{2}{c}{NS}& \multicolumn{2}{c}{SS}& \multicolumn{2}{c}{MorphZ$_{(\mathrm{NS})}$}&\multicolumn{2}{c}{MorphZ$_{(\mathrm{PT})}$}\\\midrule

Dataset &$\log(\hat{z})$&$\sigma_{\text{NS}}$& $\Delta \log(\hat{z})$&$\sigma_{\text{SS}}$& $\Delta\log(\hat{z})$&std& $\Delta\log(\hat{z})$&std\\
GW150914 &-7268.391&0.108& -4.007&0.122& -0.178&0.366& -0.246&0.415\\ \bottomrule

\end{tabular}
\end{table*}

To ensure a fair comparison across methods, we compare \morphz estimates using each sampler's posterior (identical likelihoods and priors). This results in two independent estimates: $\morphz_{\text{(NS)}}$ and $\morphz_{\text{(PT)}}$. For both estimates, a second order Morph approximation $\mathcal{M}_{2}$ is constructed using 5000 NS and 2000 PT posterior samples and \textit{Silverman}'s rule for KDE, and 5000 likelihood calls for the BS phase. \deleted{A} 100 \morphz estimates are obtained for each case (each individual estimate takes a few seconds to compute). For each dataset, we report the evidences from NS, SS, and both \morphz independent estimates. We compare methods primarily via $\Delta\log(z)$ relative to NS.

For the 100 BBH simulations we compute the evidence difference 
$\Delta \log(\hat{z}) = \log(\hat{z})_{\mathrm{method}} - \log(\hat{z})_{\mathrm{NS}}$. The $\log(z)_{\mathrm{NS}}$ is used as a reference estimate of the evidence, although it does not necessarily coincide with the true $\log(z)$ and may itself be subject to error. For NS and SS,   $\sigma_{\text{NS}} \approx \sqrt{H/n_{\mathrm{live}}}$ is nested sampling standard error\added{, where $H$ is the information gained about the parameters from the data, measured by the posterior relative to the prior~\cite{skilling2006nested},} and $\sigma_{\text{SS}}$ is the bootstrap estimated error \cite{Ashton2021}. All $|\Delta \log(\hat{z})|$ for $\morphz_{\text{(NS)}}$ and $\morphz_{\text{(PT)}}$ are shown in Figure~\ref{fig:logz-error-lvk}.
Table~\ref{tab:ensemble-deltas} summarizes \deleted{theses} \added{these} results. 
Both \morphz variants show small offsets, typically with $|\Delta \log(\hat{z})| \lesssim 0.1$. 
The population spread is also small with standard deviations around 0.22. 
More than $95\%$ of simulations satisfy $|\Delta \log(\hat{z})| < 0.5$, and the whole population reaches $|\Delta \log(\hat{z})| < 1$ enabling unbiased model comparison between signal and noise models. 
By contrast, steppingstone sampling evidences runs show a much larger scatter and a clear bias, which is consistent with expectations for SS estimates computed with a coarse temperature ladder. These results show that the number of PT \deleted{(}temperatures used to obtain the posterior samples remains sufficient for \morphz to obtain accurate evidence estimates; however, they are not enough to obtain reliable SS estimates. 

We next analyze the GW150914 event. 
Table~\ref{tab:gw1509} shows that both \morphz independent estimates differ from the NS baseline by less than $|\Delta \log(\hat{z})| \lesssim 0.25$.  
Differences at the level of $|\Delta \log(\hat{z})| \lesssim 0.2$ are comparable to
the natural run-to-run variation expected from nested sampling with
$n_{\mathrm{live}}\sim 10^3$, since the evidence uncertainty scales as
$\sigma_{\text{NS}} $ for typical
information values $H \sim 10{-}50$~\cite{skilling2006nested,Ashton2019}.
The SS estimate from the GW150914 \bilbymcmc run is clearly biased relative to the nested-sampling baseline.~\footnote{
We assess posterior consistency for GW150914 using the Jensen--Shannon divergence (JSD) computed from 50-bin one-dimensional marginals. 
The median JSDs were 0.126 for \dynesty compared with the GWTC-2.1 posterior, 0.121 for \bilbymcmc compared with GWTC-2.1, and 0.073 between our two samplers. 
The largest values (around 0.15 to 0.18) occurred for luminosity distance and $\chi_{\mathrm{p}}$. 
Crucially, these differences do not impact the goal of this work, which is accurate evidence $\log(z)$ estimation rather than precise agreement of posterior marginals.}

These results show that \morphz provides accurate and computationally efficient evidence estimates for CBC gravitational-wave analyses. Across both the simulation study and GW150914, \morphz reproduces NS evidences within $|\Delta \log(\hat{z})|$ of order 0.1 to 0.25. Most importantly, \morphz remains accurate when applied to posterior samples generated by low-temperature-resolution PT runs even though the corresponding SS evidences are biased. Because \morphz requires only posterior samples and access to the likelihood and prior, it enables reliable evidence estimation without specialized evidence-targeted runs. This makes it well suited for rapid model selection, large simulation campaigns, and retrospective analyses of existing LVK posteriors.

\section{Discussion}
\label{discuss}

\subsection{Constructing the Morph approximation}
In this work, the Morph approximation showed its consistency and efficiency across different examples using kernel density estimates and the seeded greedy maximization Algorithm~\ref{sgm}. The total correlation of all possible blocks is required to build such an approximation. For block sizes $L\geq3$, the evaluation of the total correlation over all possible blocks becomes increasingly expensive in dimensions exceeding $\approx136$ (see Section~\ref{appli:pta}). Empirically, we find that thinning the sample size to the order of a few hundred keeps the cost significantly low while yielding similar total correlation estimates to those obtained with larger sample sizes. For $L\geq3$, the total correlation can also be used to probe higher-order (linear and non-linear) dependencies among parameters, highlighting when they are physically linked by common mechanisms and potentially revealing new, otherwise  hidden couplings in the model.

\begin{figure}
    \centering
    \includegraphics[width=0.97\linewidth]{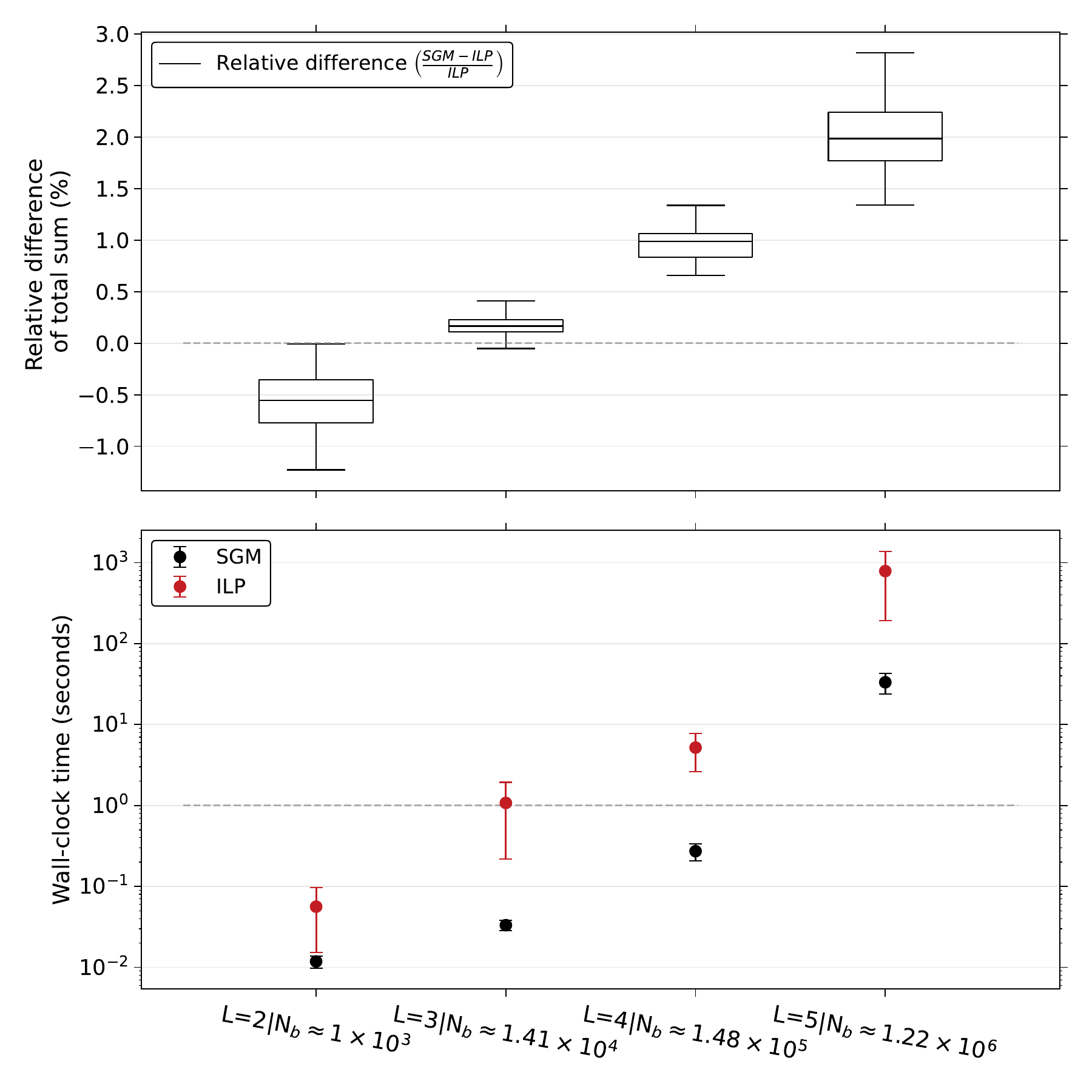}
    \caption{Relative difference of the total sum of simulated weights and runtime comparison of SGM and ILP across different orders of the Morph approximation for 100 independent sum of weight maximization simulation. Each simulation assign a weight drawn from $\text{Beta}(\alpha,2)$ with $\alpha\sim\text{Uniform}(5,9)$ for each possible group from $\binom{50}{L}$.    Top: Relative difference of the total sum of chosen weights, \(((\mathrm{SGM}-\mathrm{ILP})/\mathrm{ILP})\times 100\%)\), shown as box-and-whisker plots for $L=2-5$ with the corresponding $\text{N}_b$ number of blocks of length $L$  ; the horizontal dashed line marks zero difference. Bottom: Wall-clock time in seconds for SGM (black) and ILP (red) on the same instances; markers denote mean and (\(\pm1\sigma\)). The dashed line marks \(1\mathrm{s}\).}
  \label{fig:sgm_ilp_accuracy_runtime}
\end{figure}

While this implementation of the Morph approximation is efficient, both the SGM algorithm and KDEs have intrinsic limitations. Although an optimal Morph approximation exists in principle for $2\leq L\leq d$, it can be obtained in practice only for $L=2$, where the problem reduces to maximum-weight graph matching. This problem is solved exactly via Blossom-based algorithms \cite{edmonds1965paths,galil1986efficient}. While for $\mathcal{M}_{L>2}$, an approximately optimal solution can be acquired by solving an integer-linear-programming (ILP) set-packing formulation using, for instance, \project{PuLP}\cite{mitchell2011pulp,forrest2005cbc}, since matching-based methods no longer apply. The optimal solution problem becomes increasingly difficult. For example in Figure~\ref{fig:sgm_ilp_accuracy_runtime}, the 50-dimensional case with $L=5$, an algorithm has to select 10 blocks out of $\approx1.22\times10^{6}$ possible choices that maximizes the total sum of the 10 blocks' weights. This performance comparison, presented in Figure~\ref{fig:sgm_ilp_accuracy_runtime}, shows that SGM is significantly faster and outperforms ILP for $L\ge3$.  However, the SGM algorithm can be sensitive to the choice of initialization seed, may potentially converge to suboptimal local maxima, and its approximation error may increase for larger block sizes $L$. Since an optimal Morph approximation cannot be computed for orders three and higher, a direct quantitative comparison between different orders of the Morph approximation might not be faithful to the theoretical expectations for high dimensional problems (see example in Figure~\ref{fig:morph_Dkl}).

\added{The construction of the Morph approximation might possibly generate singletons that are correlated and therefore not well represented, since they are assumed independent by definition in \mbox{Equation~\ref{eq:M_b}}. By default, the algorithm defines singletons as the parameters with the lowest multi-information, which may partially mitigate this under-representation. However, the quality of the marginal likelihood estimate is not significantly affected by this issue (see \mbox{Figure~\ref{fig:morph_Dkl}).} Nonetheless, several potential approaches could be considered to account for correlations among singletons. For instance, one could model the singletons as one block of length $\leq L-1$ rather than a product of independent marginals.}

In parallel, the performance of KDEs degrades rapidly with \added{higher} dimension\added{s}, where naïve KDEs become unreliable beyond $d\geq10$ dimensions \cite{Tsybakov2009,virtanen2020scipy}. Although \textit{Silverman}'s rule of thumb often suffices, KDE bandwidth choice is critical and largely determines the bias–variance balance; alternative methods can be used such as the Improved Sheather–Jones bandwidth~\cite{Botev2010DiffusionKDE} or bandwidth selection through isotropic cross-validation~\cite{silverman2018density,duong2005cross}. 

\subsection{Evidence estimation using \morphz }

For marginal likelihood estimation, the Morph approximation serves as an importance sampling distribution for evidence estimators, and is especially effective for bridge sampling.  We tested the Morph approximation on importance sampling (IS) estimator \cite{chan2015marginal} for the models presented in Table \ref{tab:pta} and found that the estimator may occasionally be unstable for high-dimensional models (results not shown).  IS requires an importance density with heavier tails than the posterior to avoid instability. In contrast, bridge sampling keeps the ratios bounded under both thinner and heavier tails mismatches provided proposal and posterior overlap. Beyond tail behavior, bridge sampling places fewer constraints on the proposal distribution making it easier to construct suitable proposals. Using the Morph approximation together with the optimal bridge function and the iterative scheme, bridge sampling becomes nearly an automatic flow to implement and obtain accurate, efficient marginal-likelihood estimates, even for hierarchical models. This implementation reaches its full potential when posterior coverage is adequate, allowing the evidence to be obtained with only a few additional likelihood evaluations. For reliable evidence estimation, \morphz requires at least minimal information about the posterior beyond the burn-in phase. The factored structure of the Morph approximation enables immediate generation of independent proposal draws in the current implementation. When posterior coverage is sparse, the newly accepted samples can be merged with previously collected samples to complete the posterior coverage. 

\section{Conclusion}
\label{conclusion}

In this work, we showed that \morphz can deliver accurate evidence at substantially lower cost across statistical problems, real PTA noise and GWB inference, and LVK-CBC applications, spanning 2–136 parameters. 

On statistical benchmarks, it matches or exceeds nested sampling accuracy while cutting likelihood calls by $\sim2$ orders of magnitude; uniquely, it remains accurate on the challenging peak-plateau problem where path-based estimators struggle and standard NS requires more sampling efforts. \morphz also improves NS estimates even when posterior coverage is still incomplete. In PTA models, it attains precise evidence with $\sim10–20$ calls for $d$=8–16, $\sim300–500$ for $d$=32–74, and $\sim20\times$ fewer calls than GSS at $d$=136 even for expensive likelihoods, with bridge sampling relative errors that conservatively track empirical ones. In CBC analyses, \morphz reproduces NS evidence values with $|\Delta\log (\hat{z})|\leq0.1$ on simulations and $|\Delta\log (\hat{z})|\leq0.25$ for GW150914, while SS requires a greater number of temperatures for a more accurate estimate.

Currently, \morphz can complement existing samplers by acting as a lightweight, post-processing evidence engine that needs only posterior draws. It can be paired with an NS run at reduced live points to explore and sample, then used to refine evidence accurately at far lower cost; likewise, it remains reliable when fed posterior samples from coarse-temperature PT-MCMC runs that are insufficient for unbiased SS estimates, enabling evidence estimation without dedicated evidence-targeted runs and facilitating retrospective analyses. Moreover, it can leverage posterior samples from fast, gradient-based samplers (e.g., HMC \cite{Duane1987HMC}, NUTS \cite{HoffmanGelman2014NUTS}, and Riemannian variants \cite{GirolamiCalderhead2011RMHMC}), turning their high-quality draws into accurate evidence estimates with minimal additional computation.

Looking ahead, the standard KDE used in Morph approximation construction, although efficient, could be replaced by richer models, such as copulas or normalizing flows, that better capture dependence and tail behavior, reducing bias and variance at a given sample size. Future work will also parallelize likelihood evaluation (vectorized and batched across bridges and sample blocks) to cut wall-clock time without increasing the number of calls, pushing \morphz’s efficiency to new frontiers. Finally, bootstrap methods (parametric and nonparametric) will be explored to refine the relative-error estimates of bridge sampling and make the reported uncertainties track empirical variability more closely.

\begin{acknowledgments}
EMZ, PMR, AV, WS, and RM gratefully acknowledge support by the Marsden Fund Council grant MFP-UOA2131 from New Zealand Government funding, managed by the Royal Society Te Apārangi. 
This work was performed on the OzSTAR national facility at Swinburne University of Technology. The OzSTAR program receives funding in part from the Astronomy National Collaborative Research Infrastructure Strategy (NCRIS) allocation provided by the Australian Government, and from the Victorian Higher Education State Investment Fund (VHESIF) provided by the Victorian Government.\\
\textit{Software}: \texttt{ENTERPRISE} \cite{ellis2019enterprise,enterprise}, \texttt{enterprise\_extension} \cite{enterprise}, \cite{justin_ellis_2017_1037579}  ,\cite{dalcin2005mpi}, \texttt{Jupyter} \cite{kluyver2016jupyter}, \texttt{matplotlib} \cite{hunter2007matplotlib}, \texttt{numpy} \cite{harris2020array}, \texttt{scipy} \cite{pauli_virtanen_2020_4406806}, \texttt{arviz} \cite{kumar2019arviz}.

\end{acknowledgments}

\section*{Data and Software Availability}
\label{data}
The \morphz estimator is available as the open-source MorphZ Python package on GitHub
(\url{https://github.com/EL-MZ/MorphZ}) and on PyPI (\url{https://pypi.org/project/morphZ/})
under the BSD 3-Clause License.
\added{The code used for the analysis, along with links to the public datasets, can be found on GitHub \mbox{(\url{https://github.com/EL-MZ/MorphZ})} }

\deleted{The data products generated and analyzed in this study will be archived on Zenodo and made publicly
available. The corresponding Zenodo DOI will be included in this section in the final published
version of the manuscript.}

\appendix
\section{KL divergence between $P$ and $\mathcal{M}_{\mathcal{B}_{L}}$ in terms of total correlation}
\label{app:kL TC derivation}

We recall the entropies
\begin{align}
H(\boldsymbol{\theta}) &= -\mathbb{E}_P[\log P(\boldsymbol{\theta})],\\
H(\boldsymbol{\theta}_b) &= -\mathbb{E}_{P_b}[\log P_b(\boldsymbol{\theta}_b)],\\
H(\theta_s) &= -\mathbb{E}_{P_s}[\log P_s(\theta_s)],
\end{align}
and define the block total correlation by
\begin{equation}
\mathcal{C}(\boldsymbol{\theta}_b) = \sum_{i\in b} H(\theta_i) - H(\boldsymbol{\theta}_b).
\end{equation}

We now derive an expression for $D_{\mathrm{KL}}\!\big(P\|\mathcal{M}_{\mathcal{B}_{L}}\big)$ in terms of the total correlation. The \textit{Kullback–Leibler} between $P$, the probability density, and $\mathcal{M}_{\mathcal{B}_{L}}$, the Morph approximation of order $L$, is given by
\begin{align}
&D_{\mathrm{KL}}\!\big(P\|\mathcal{M}_{\mathcal{B}_{L}}\big)
= \mathbb{E}_P\!\left[\log P(\boldsymbol{\theta})-\log \mathcal{M}_{\mathcal{B}_{L}}(\boldsymbol{\theta})\right]\\
&= \mathbb{E}_P\!\left[\log P(\boldsymbol{\theta})-
\sum_{b\in\mathcal B}\log P_b(\boldsymbol{\theta}_b)-
\sum_{s\in \mathcal{S}}\log P_s(\theta_s)\right] \\
&= -H(\theta)
   +\sum_{b\in\mathcal B}\big(-\mathbb{E}_P[\log P_b(\boldsymbol{\theta}_b)]\big)\\
   &\quad+\sum_{s\in \mathcal{S}}\big(-\mathbb{E}_P[\log P_s(\theta_s)]\big).
\end{align}
Assuming that $P_b$ and $P_s$ denote the corresponding marginals of $P$ (that is,
$P_b$ is the joint marginal distribution of the block $b$ and $P_s$ is the marginal
distribution of the singleton $s$), we have
\begin{align}
    \mathbb{E}_P[\log P_b(\boldsymbol{\theta}_b)] = \mathbb{E}_{P_b}[\log P_b(\boldsymbol{\theta}_b)],
\end{align}
and similarly for $P_s$ we have
\begin{align}
    \mathbb{E}_P[\log P_s(\theta_s)] = \mathbb{E}_{P_s}[\log P_s(\theta_s)].
\end{align}
Hence,
\begin{align}
&D_{\mathrm{KL}}\!\big(P\|\mathcal{M}_{\mathcal{B}_{L}}\big)
= -H(\boldsymbol{\theta})+\sum_{b\in\mathcal B} H(\boldsymbol{\theta}_b)+\sum_{s\in \mathcal{S}} H(\theta_s)
\\
&\left(\text{since $\sum_{i\in\Gamma}H(\theta_{i})
  =\sum_{s\in \mathcal{S}} H(\theta_s)+\sum_{b\in\mathcal B}\sum_{i\in b} H(\theta_i)$}\right) \\
&=\Big(\sum_{i\in\Gamma} H(\theta_i)-H(\boldsymbol{\theta})\Big)
 -\sum_{b\in\mathcal B}\Big(\sum_{i\in b} H(\theta_i)-H(\boldsymbol{\theta}_b)\Big) \\
&=\Big(\sum_{i\in\Gamma} H(\theta_i)-H(\boldsymbol{\theta})\Big)
 -\sum_{b\in\mathcal B}\mathcal{C}(\boldsymbol{\theta}_b).
\end{align}
Equivalently,
\begin{align}
&D_{\mathrm{KL}}(P\|\mathcal{M}_{\mathcal{B}_{L}})
= \mathcal{C}(\boldsymbol{\theta})-\sum_{b\in\mathcal B}\mathcal{C}(\boldsymbol{\theta}_b),
\end{align}
where $\mathcal{C}(\theta)=\sum_{i\in\Gamma}H(\theta_i)-H(\boldsymbol{\theta})$ is the total correlation
of the full vector, and $\Gamma$ is the index set of all variables.
All quantities are assumed finite.
\newline

\bibliographystyle{apsrev4-2}

\bibliography{biblio}

\end{document}